\documentclass{sf2a-conf2015}
\usepackage{graphicx}
\usepackage{hyperref}
\usepackage[]{natbib}  
\usepackage{epstopdf}

\def\BibTeX{{\rm B\kern-.05em{\sc i\kern-.025em b}\kern-.08em
    T\kern-.1667em\lower.7ex\hbox{E}\kern-.125emX}}
\bibpunct{(}{)}{;}{a}{}{,}  

\begin{document}

\TitreGlobal{SF2A 2015}


\title{The ``Binarity and Magnetic Interactions in various classes of stars" (BinaMIcS) project}

\runningtitle{BinaMIcS}

\author{C. Neiner}\address{LESIA, Observatoire de Paris, PSL Research
University, CNRS, Sorbonne Universit\'es, UPMC Univ. Paris 06, Univ. Paris
Diderot, Sorbonne Paris Cit\'e, 5 place Jules Janssen, 92195 Meudon, France}

\author{J. Morin}\address{LUPM-UMR 5299, CNRS \& Universit\'e Montpellier, Place
Eug\`ene Bataillon, 34095, Montpellier Cedex 05, France}

\author{E. Alecian$^{1,}$}\address{Univ. Grenoble Alpes, CNRS, IPAG, F-38000
Grenoble, France}

\author{the BinaMIcS collaboration}

\setcounter{page}{237}


\maketitle


\begin{abstract}
The ``Binarity and Magnetic Interactions in various classes of stars" (BinaMIcS)
project is based on two large programs of spectropolarimetric observations with
ESPaDOnS at CFHT and Narval at TBL. Three samples of spectroscopic binaries with
two spectra (SB2) are observed: known cool magnetic binaries, the few known hot
magnetic binaries, and a survey sample of hot binaries to search for additional
hot magnetic binaries. The goal of BinaMIcS is to understand the complex
interplay between stellar magnetism and binarity. To this aim, we will
characterise and model the magnetic fields, magnetospheric structure and
coupling of both components of hot and cool close binary systems over a
significant range of evolutionary stages, to confront current theories and
trigger new ones. First results already provided interesting clues, e.g. about
the origin of magnetism in hot stars. 
\end{abstract}



\section{Introduction}

The goals of the "Binarity and Magnetic Interactions in various classes of
stars" (BinaMIcS) project are to understand the impact of magnetic fields on
stellar formation and evolution, of tidal effects on fossil and dynamo magnetic
fields, of magnetism on angular momemtum and mass transfers between binary
components, as well as magnetospheric interactions. To address these questions,
we are missing observational information on the magnetic field strengths and
topologies of a statistically large sample of magnetic close binary systems, in
which we expect significant interaction via tidal or magnetospheric
interactions. Therefore, BinaMIcS is based on two large programs of
spectropolarimetric observations with ESPaDOnS at CFHT (Hawaii) and Narval at
TBL (Pic du Midi, France), in addition to theoretical developments and modelling
efforts.\\ 

Three samples of short-period spectroscopic binaries with two spectra (SB2) are observed:
\begin{itemize}
\item known selected cool ($<$ F5) magnetic binaries, to characterise their magnetic properties in details and compare them to single magnetic cool stars
\item the few known hot ($>$ F5) magnetic binaries, to characterise their magnetic properties in details and compare them to single magnetic hot stars
\item a survey sample of hot binaries, to discover new hot magnetic binaries and compare the occurence of magnetic fields in hot binaries versus single hot stars.
\end{itemize}

\section{Magnetic hot SB2 systems}

Six magnetic SB2 systems with at least one O, B, or A component were known to
exist and are currently being characterized in the frame of BinaMIcS. These are 
HD\,37017 \citep[V1046\,Ori,][]{bohlender1987}, HD\,37061
\citep[NU\,Ori,][]{petit2011}, HD\,136504
\citep[$\epsilon$\,Lup,][]{hubrig2011,shultz2012}, HD\,47129 \citep[Plaskett's
star,][]{grunhut2013}, HD\,98088 \citep{folsom2013}, and HD\,5550 (Neiner et
al., these proceedings, Alecian et al., submitted). Among them, only one is
known to host two magnetic stars: $\epsilon$\,Lup \citep{shultz2015}. In
addition, two newly discovered magnetic SB2 systems with early F stars are being
characterized within BinaMIcS: HD\,160922 \citep{neiner2013} and HD\,210027
\citep[$\iota$\,Peg,][]{neiner2014}. 

No other magnetic OBA SB2 system was discovered among the $\sim$200 systems
($\sim$400 stars) observed within the survey sample. Among single hot stars,
$\sim$7\% are found to host a magnetic field, with a typical strength of a few
hundreds to a few thousands gauss \citep{grunhut2015}. These fields are of
simple configuration and stable over decades. Moreover, they are of fossil
origin, i.e. remnants from the field present in the molecular cloud at the time
of stellar formation, possibly enhanced by a dynamo action during the early
phases of the life of the star \citep{borra1982,neiner2015}. If
similar magnetic fields were present with the same occurence in hot binaries as
in single hot stars, we should have detected 20 to 30 magnetic stars in the
survey sample. There is thus a clear and strong deficit of magnetic stars in
hot short-period spectroscopic binaries. 

A possible explanation for this dearth of magnetic fields in hot SB2 systems is
provided by stellar formation processes. Stellar formation simulations
\citep[e.g.][]{commercon2011} showed that fragmentation of dense stellar cores
is inhibited when the medium is magnetic. Therefore, it seems that it is more
difficult to form a binary system in the presence of a fossil field and, thus,
magnetic hot binaries are rarer. 

\section{Magnetic cool SB2 systems}

\begin{figure}[t!]
 \centering
 \includegraphics[width=0.32\textwidth,clip]{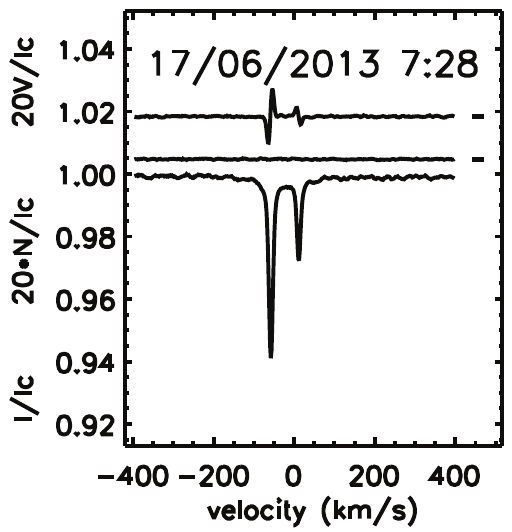}      
 \includegraphics[width=0.32\textwidth,clip]{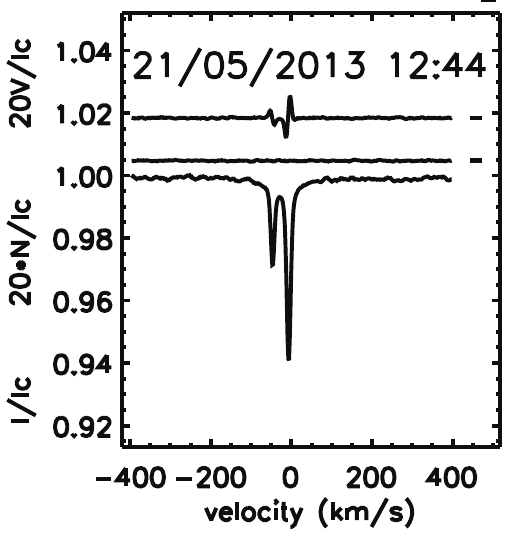}      
 \includegraphics[width=0.32\textwidth,clip]{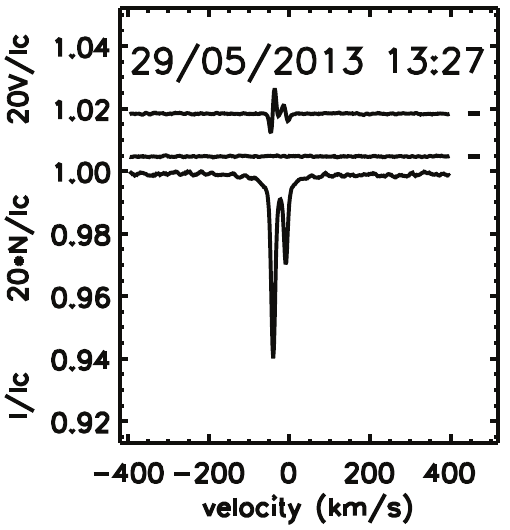}      
  \caption{Three examples of LSD Stokes V (top), null (middle), and intensity profiles (bottom) of the cool binary system BY\,Dra, taken at various orbital phases. The V and N profiles have been shifted upwards for display purposes. The Stokes V profiles clearly show that both binary components are magnetic.}
  \label{Neiner2:fig1}
\end{figure}

The BinaMIcS sample of magnetic cool SB2 systems includes cool main sequence
stars, evolved RS CVn objects, and young T Tauri stars. Several of these systems
show signatures of magnetic fields in both components. This is the case, for
example, of the K4Ve+K7.5Ve system BY Dra (see Fig.~\ref{Neiner2:fig1}) or
$\sigma^2$\,CrB \citep[see][]{neiner2013}.

\begin{figure}[t!]
 \centering
 \includegraphics[width=0.8\textwidth,clip]{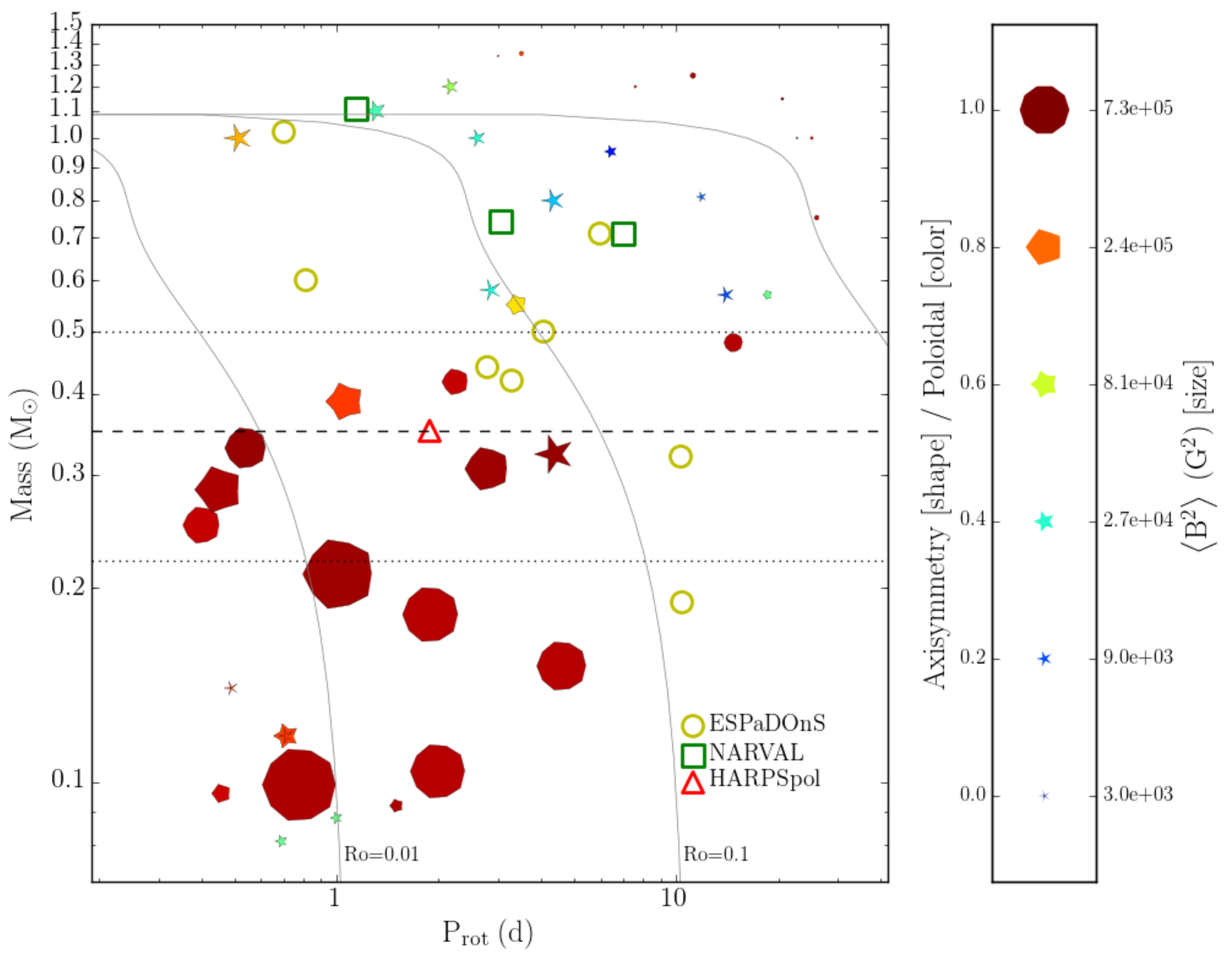}      
  \caption{Mass versus rotation period diagram for single magnetic stars (filled
symbols) and cool binaries of the BinaMIcS sample (open symbols). For single
stars, the shape of the symbol indicates the axisymmetry of the magnetic field
configuration, its color indicates how poloidal the field is, and its size 
indicates the strength of the field.}
  \label{Neiner2:fig2}
\end{figure}

Single magnetic cool stars show various types of magnetic fields, with various
levels of complexity, axisymmetry, and strength. Fig.~\ref{Neiner2:fig2} shows
known single magnetic cool stars (filled symbols) in a mass versus rotation
period diagram. Stars with the strongest fields are the lower mass stars, the
slowest rotators, and often have simple poloidal fields. Weak complex fields are
found in two regions of the diagram: slowly rotating very low-mass stars, and
rapidly rotating higher mass stars. Open symbols in Fig.~\ref{Neiner2:fig2}
indicate the position of the cool magnetic binary targets of BinaMIcS. Magnetic
maps of these binaries will be compared to the maps of single stars to assess
the effect of binarity on stellar dynamo processes.

\section{Conclusions}

BinaMIcS aims at studying the effect of magnetism on binary formation and
evolution. Individual hot and cool short-period spectroscopic binary (SB2)
systems are being studied in great details to infer their magnetic properties.
These results will be compared to those obtained for single stars. In addition,
the magnetic survey of hot binaries already provided an important result: hot
stars in short-period binaries are less often magnetic than when they are
single. 

\begin{acknowledgements}
We thank the ``Programme National de Physique Stellaire" (PNPS) of CNRS/INSU
(France) for their financial support to the BinaMIcS project.
\end{acknowledgements}

\bibliographystyle{aa}  
\bibliography{Neiner2} 

\end{document}